\documentclass[fleqn,usenatbib]{mnras}

\usepackage{newtxtext,newtxmath}
\usepackage[T1]{fontenc}

\DeclareRobustCommand{\VAN}[3]{#2}
\let\VANthebibliography\thebibliography
\def\thebibliography{\DeclareRobustCommand{\VAN}[3]{##3}\VANthebibliography}

\usepackage{subcaption}
\usepackage{multicol}
\usepackage{graphicx}	% Including figure files
\usepackage{amsmath}	% Advanced maths commands
\usepackage{xcolor}
\usepackage{float}
\usepackage{afterpage}
\usepackage{caption}
\title[Complexity Variability in Seyfert AGN]{Complexity and Multifractal Variability in Multi-Band Emission of Seyfert AGN}

\author[R. A. A. Souza et al.]{
 R. A. A. Souza$^{1,2}$, E. Hatziminaoglou$^{3,4,5}$, A. de Pádua Santos$^{2}$ ,T. Stosic$^{6}$\\ 
$^{1}$Physics Program, Universidade de Pernambuco, Campus Mata Norte, Nazar\'{e} da Mata, PE 55800-000, Brazil \\
$^{2}$Departamento de Física, Universidade Federal Rural de Pernambuco, Recife, PE, 50670-901, Brazil \\
$^{3}$Instituto de Astrof\'{i}sica de Canarias, 38205 La Laguna, Tenerife, Spain \\
$^{4}$ESO, Karl-Schwarzschild-Str. 2, 85748 Garching bei M\"unchen, Germany \\
$^{5}$Departamento de Astrof\'{i}sica, Universidad de La Laguna, 38206 La Laguna, Tenerife, Spain\\
$^{6}$Departamento de Estatística e Informática, Universidade Federal Rural de Pernambuco, Recife, PE, 50670-901, Brazil
}
\date{Accepted 2026 May 06. Received 2026 January 14; in original form 2025 July 04.}

\pubyear{2026}

\begin{document}
\label{firstpage}
\pagerange{\pageref{firstpage}--\pageref{lastpage}}
\maketitle

\begin{abstract}
Active galactic nuclei (AGNs) exhibit complex variability across multiple wavelengths, reflecting diverse physical processes near their central engines. This work investigates the temporal variability of four AGNs Mrk~509, NGC~5548, NGC~4151, and NGC~4593 using multifractal detrended moving average (MFDMA) analysis and Fisher-Shannon information plane applied to their X-ray, ultraviolet, and optical light curves. These methods quantify the scaling behavior and complexity of the variability, revealing persistent correlations and distinct variability patterns across energy bands. The Fisher-Shannon analysis further characterizes the degree of stochasticity and structural complexity in the emission processes. Our findings support the interpretation that multifractal and information-theoretic measures provide effective diagnostics of the physical mechanisms driving AGN variability. This study demonstrates the utility of advanced time series techniques as effective diagnostics of AGN variability mechanisms.
\end{abstract}


%%%%%%%%%%%%%%%%%%%%%%%%%%%%%%%%%%%%%%%%%%%%%%%%%%

%%%%%%%%%%%%%%%%% BODY OF PAPER %%%%%%%%%%%%%%%%%%

\section{Introduction}\label{sec:intro}

AGN variability presents a persistent challenge due to its inherently nonlinear and irregular nature. Light curves often display asymmetric and source-dependent features, complicating efforts to identify universal variability patterns \citep{peterson2000x,vaughan2003characterizing}. Indeed, quasar variability emerges as a fundamentally stochastic phenomenon, where each light curve represents a single realization of a random process. Consequently, the variability measured in an AGN can be influenced by multiple factors. The relative contribution of physical processes intrinsic to the source may change with wavelength, and external effects, such as the intergalactic environment, may introduce additional sources of variability, for example gravitational microlensing. As a result, each observational epoch of a quasar represents a unique realization of its variability, and statistical effects become relevant when the measured variability no longer faithfully reflects the energy band under investigation, as discussed by \citet{vaughan2003characterizing}. Traditional time series analyses, typically suited for linear and Gaussian processes \citep{Vio1992}, are inadequate to fully capture the complexity of active galactic nuclei (AGN) variability. Given that quasars are dynamic systems likely governed by nonlinear and chaotic mechanisms \citep{Chian1997}, it becomes necessary to employ analysis techniques capable of revealing such intrinsic nonlinearity. Notably, optical quasar light curves have been shown to arise from nonlinear and non-stationary stochastic processes \citep{Longo1996,Green1999,Provenzale1994}. The presence of multifractality in these signals suggests intermittent and scale-dependent variability consistent with complex physical processes, but it does not by itself constitute definitive evidence for intrinsically nonlinear dynamical mechanisms, as similar signatures may also arise from correlated stochastic processes \citep{Longo1996,belete2018multifractality,belete2019revealing,2019MNRAS.488.3274D,2021A&A...650A..40D}.

 In response to these challenges, recent studies have adopted multifractal analysis to better characterize AGN variability. Notably, \citet{assis2024multifractality} investigated the multifractal behavior of 14 gravitationally lensed quasars using the Wavelet Transform Modulus Maxima (WTMM) method, confirming the presence of nonlinear signatures in all systems. These results support the interpretation of AGN variability as intrinsically multifractal, shaped by both source-intrinsic and lensing-induced effects. Complementary to this, \citet{belete2019novel} applied the MFDMA method to the light curves of NGC~5548, revealing a nonlinear relationship between the 5100 Å continuum and the H$\beta$ emission line. Their findings suggest that the broad-line region (BLR) reprocesses radiation from the central engine in a complex, nonlinear fashion, as evidenced by the stronger multifractality in the H$\beta$ line compared to the continuum. These results highlight the importance of nonlinear time series analysis in probing the physical mechanisms underlying AGN variability, such as radiation reprocessing, reverberation, and temporal correlations across emission regions. Similar methodologies have also been employed in the study of other AGNs, including 3C 273 \citep{belete2018multifractality} and Q0957+561 \citep{belete2019revealing}.

In parallel with the previous discussion on multifractality in complex signals, a particularly analytical tool emerges from the combination of Shannon entropy \citep{shannon1948} and Fisher information \citep{fisher1925}: the Fisher–Shannon information plane. This bivariate representation allows for the simultaneous evaluation of global uncertainty - captured by Shannon entropy - and local order, as quantified by Fisher information. Numerous studies across diverse fields have demonstrated the discriminative power of this approach. It has been employed to analyze the presence of heavy metals in atmospheric particulate matter \citep{telesca2009cd}, wind speed records \citep{telesca2011wind, telesca2013wind}, magnetotelluric data \citep{telesca2011magneto}, as well as microseismic tremors \citep{telesca2015micro} and volcanic tremors \citep{telesca2010stromboli}. In seismology, this method has proven effective in distinguishing between tsunamigenic and non-tsunamigenic earthquakes \citep{telesca2013tsunami, telesca2015tsunami}. Its robustness and versatility make it a valuable tool for characterizing nonlinear and multifractal dynamics. In astrophysics, for instance, \citet{lovallo2011complexity} applied Fisher information, Shannon entropy, and statistical disequilibrium to daily X-ray light curves from various sources across different energy bands. Their results revealed a universal pattern of complex behavior, independent of the source and energy range, suggesting that the intrinsic variability of astrophysical signals carries shared features that are effectively captured by the Fisher–Shannon information plane. 

This paper is organized as follows. Section \ref{sec:observationaldata} presents the observational data, including a description of the AGN sample variability and details on the monitoring strategy and data reduction. Section \ref{sec:method} outlines the methodology, covering the Multifractal Detrended Moving Average (MFDMA) technique, the Fisher-Shannon entropy plane, and the complexity quantifier. In Section \ref{sec:results}, we discuss the results obtained from the multifractal and Fisher-Shannon analyses and explore whether complexity measures can provide new insights into quasar variability. Finally, Section \ref{sec:conclusion} summarizes the main conclusions of this work.

\section{Observational Data}
\label{sec:observationaldata}

\begin{figure*}
    \centering
    \includegraphics[width=13cm]{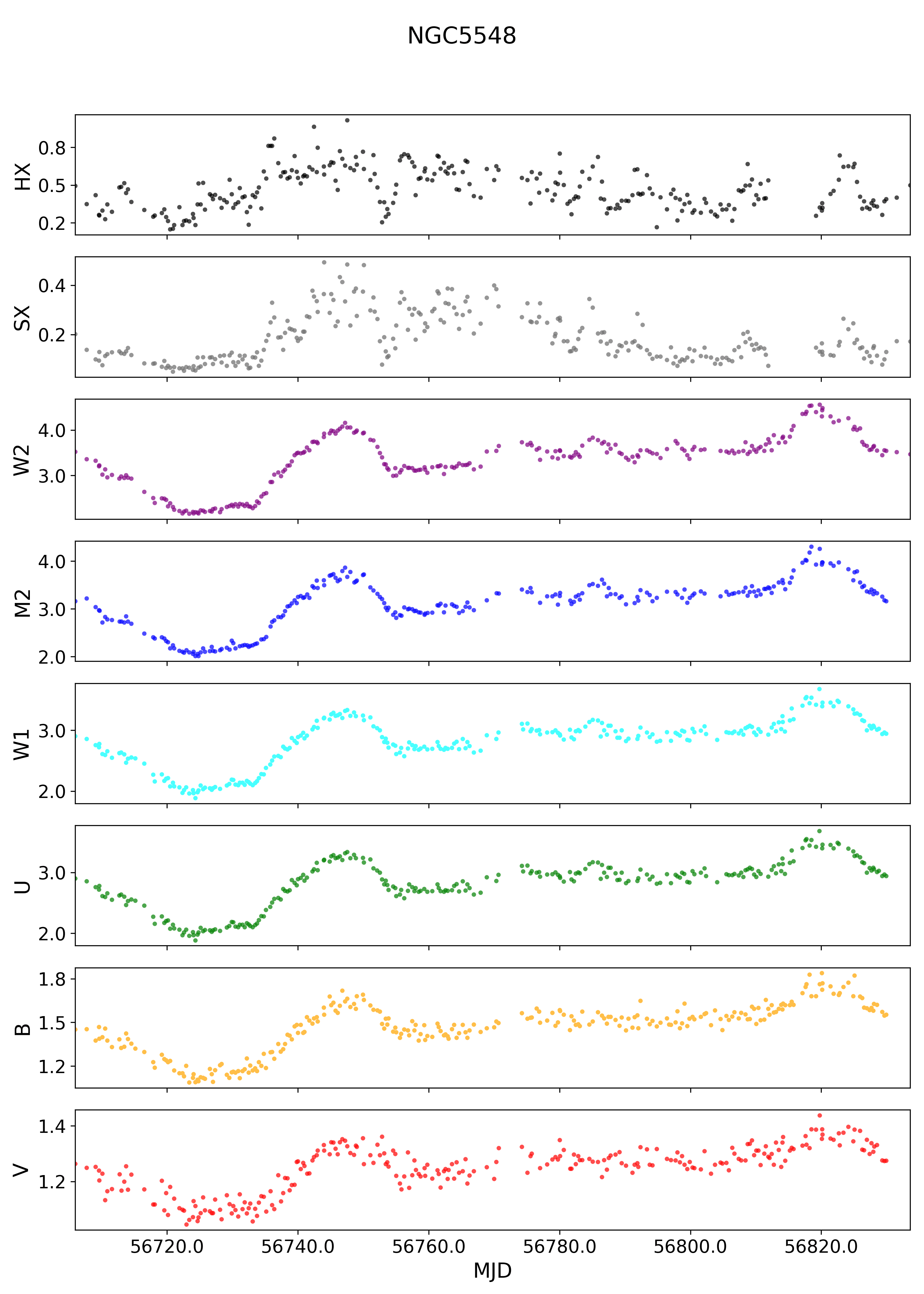}
      \caption{The original light curve of NGC~5548 for all eight observational bands. Light curves of the analyzed AGN, ordered by wavelength from top to bottom. The top two panels correspond to XRT observations in the HX (1.5--10 keV) and SX (0.3--1.5 keV) bands, while the lower six panels show the UVOT bands (W2, M2, W1, U, B, and V). Each panel displays the flux as a function of MJD, sharing a common time axis but with independent flux scales. X-ray fluxes are given in units of counts s$^{-1}$, and UVOT fluxes are expressed in units of $10^{-14}$\,erg\,cm$^{-2}$\,s$^{-1}$\,\AA$^{-1}$.}

    \label{fig:lc_plot}
\end{figure*}

This study focuses on the variability behavior of four well-known AGNs: Mrk 509, NGC~5548, NGC~4151, and NGC~4593, each extensively observed, in a single epoch, across eight energy bands. The data analyzed in this work were collected as part of the monitoring campaign described in \citet{edelson2019first}, using the \textit{Infrared Detected Reverberation Mapping} (IDRM) technique. The sample was selected based on three main criteria. First, the light curves were obtained from a publicly available, peer-reviewed dataset that underwent a uniform data-reduction procedure, ensuring consistent time-series analyses within the community \citep{edelson2019first}. An example of these light curves is shown for NGC 5548 in Fig.~\ref{fig:lc_plot}. Second, each light curve contains more than 150 data points, meeting the minimum requirement for the reliable application of the MFDMA method, which will be described below. Third, all selected AGNs have light curves observed in the same energy bands, minimizing the likelihood of spurious results and enabling consistent comparison of physical parameters across sources in the search for common patterns.

The multiwavelength variability observed in these AGNs reveals distinct physical mechanisms responsible for emission changes. In the Seyfert 1 galaxy Mrk 509 \citep{phillips1983outflow}, a direct correlation between soft X-ray and UV flux variations was observed. This suggests that the variability is intrinsic to the AGN and modulated by changes in the spectral energy distribution (SED) incident on the surrounding gas \citep{kaastra2011}. The archetypal Seyfert 1 galaxy NGC~5548 is one of the most extensively studied nearby active galaxies, having been among the 12 original objects classified in the seminal work of \citet{seyfert1979nuclear}. Since the 1960s, it has been the focus of numerous AGN studies \citep{mehdipour2015}. Rapid variations in its X-ray spectrum have been attributed to changes in the ionization state of intervening absorbers, while significant variability in infrared coronal lines further supports the presence of an X-ray–heated wind as a key mechanism in the energy transport processes of its nuclear region \citep{mehdipour2015,landt2015}.

For NGC~4151 Seyfert 1 galaxy \citep{zoghbi2019revisiting}, long-term observations revealed asymmetric profiles in the Balmer lines, suggesting that the broad-line region (BLR) is composed of multiple kinematically distinct components \citep{shapovalova2010}. In contrast, no significant variability was detected in the near-infrared molecular emission lines, indicating excitation driven primarily by shocks or star formation rather than by the AGN itself \citep{veilleux1991}. Additionally, an analysis of the historical B-band light curve of NGC~4151 revealed a clear nonlinear and intermittent behavior, supporting the presence of complex underlying mechanisms driving its variability \citep{Longo1996}. The Seyfert 1 galaxy NGC~4593 \citep{cackett2018accretion} exhibited UV variability over a ten-month period, which was linked to fluctuations in the intrinsic AGN continuum \citep{ward1983}. Simultaneous monitoring with the \textit{Swift} satellite revealed correlated variability in the X-ray, UV, and optical bands, consistent with X-ray reprocessing by extended regions of the accretion disk or the broad-line region (BLR) \citep{mchardy2024}.

The IDRM method enhances earlier reverberation mapping campaigns by employing six UV/optical filters, increasing the sampling cadence, and significantly expanding the number of observations. The measurements are expressed in units of $10^{-14}$ erg s$^{-1}$ cm$^{-2}$ Å$^{-1}$. This approach improves the detection of short inter-band lags, enabling the investigation of signal propagation across accretion disks. The strategy was initially applied to the AGN NGC~5548 and subsequently extended to NGC~4151 \citep{edelson2015space}, NGC~4593 \citep{mchardy2018x}, and Mrk~509 \citep{edelson2019first}, ensuring simultaneous monitoring with the \textit{Hubble Space Telescope} (HST) and ground-based observatories.
UVOT data reduction followed procedures established in previous studies \citep{edelson2015space, edelson2017swift}, including refined photometric measurements, filtering of low-quality data, and application of instrumental corrections. After rigorous quality control, 6,358 exposures were validated for analysis. For the XRT data, standard instrumental corrections were applied \citep{evans2007online, evans2009methods}, and light curves were extracted in both hard and soft X-ray bands, excluding the WT observation mode to ensure uniformity in temporal analysis. The units for both X-ray bands are reported in $ct~s^{-1}$.

The resulting light curves exhibit strong variability across all UVOT bands and the hard X-rays, with evidence of variability in the soft X-rays. The similarity among the UV/optical light curves suggests that the high sampling rate of the IDRM technique is sufficient to capture their variations, whereas X-ray fluctuations occur on shorter timescales and may be undersampled - see \cite{edelson2019first} for further details. The fundamental properties of the selected AGNs, including redshift ($z$), black hole mass ($\log M_{\text{BH}}$), and Eddington ratio ($\dot{m}$) and cadence of each observation are listed in Table~\ref{tab:agn_properties}.

\begin{table*}
	\centering
	\caption{Fundamental properties of the selected AGNs, including redshift ($z$), black hole mass ($\log M_{\mathrm{BH}}$), Eddington ratio ($\dot{m}_{\mathrm{Edd}}$), campaign time span, duration, number of visits, and average sampling cadence. Redshifts are taken from the SIMBAD database. Black hole masses are adopted from the AGN Black Hole Mass Database \citep{2015BentzBlackHole}. The Eddington ratio is defined as $\dot{m}_{\mathrm{Edd}} = L_{\mathrm{bol}} / L_{\mathrm{Edd}}$. The campaign time span is given in Modified Julian Date (MJD), the duration is expressed in days, the number of visits corresponds to the total number of observations with usable data in at least one band, and the average sampling cadence is computed from the campaign duration and the number of visits.}

	\label{tab:agn_properties}
	\begin{tabular}{lcccccc}
		\hline
		Name & $z$ & $\log M_{\text{BH}}$ & $\dot{m}$ & Visits & Duration (days) & Cadence (days)\\
		\hline
		Mrk~509  & 0.0341 & 8.05 & 5\% & 257 & 272.6 & 1.065\\
		NGC~5548 & 0.0163 & 7.72 & 5\% & 291 & 127.6 & 0.440\\
		NGC~4151 & 0.0032 & 7.56 & 1\% & 322 & 69.3  & 0.216\\
		NGC~4593 & 0.0083 & 6.88 & 8\% & 194 & 22.6  & 0.117\\
		\hline
	\end{tabular}
\end{table*}

\section{Methods}\label{sec:method}

\subsection{Multifractal Detrended Moving Average}

The Multifractal Detrended Moving Average (MFDMA), introduced by \cite{gu2010detrending}, refines the Multifractal Detrended Fluctuation Analysis  \citep[MFDFA;][]{kantelhardt2002multifractal}, which itself extends the Detrended Fluctuation Analysis 
\citep[DFA;][]{peng1994mosaic}. DFA analyzes long-range correlations in time series by removing trends using polynomial fits, producing a single scaling exponent, while MFDFA examines  long-term correlations in multifractal time series,  resulting in a hierarchy of scaling exponents.  MFDMA improves upon MFDFA by removing trends through moving averages (MA), offering a more efficient approach for multifractal analysis \citep{gu2010detrending,belete2019novel}. The MFDMA algorithm is outlined as follows:

For a time series $x(t)$, the cumulative profile $y(t)$ is defined as:  
\begin{equation}
    y(t) = \sum_{i=1}^{t} x(i).
\end{equation}

The moving average $\tilde{y}(t)$ is calculated over a window of size $n$ as:
\begin{equation}
    \tilde{y}(t) = \frac{1}{n} \sum_{k=-\lfloor(n-1)\theta\rfloor}^{\lceil(n-1)(1-\theta)\rceil} y(t - k),
\end{equation}

where $\theta$ is the window position parameter. In this study, we chose $\theta = 0$ corresponding to a backward moving average. The detrended fluctuation $\epsilon(i)$ is computed as:
\begin{equation}
    \epsilon(i) = y(i) - \tilde{y}(i).
\end{equation}

The residual series $\epsilon(i)$ is divided into $N_n$ non-overlapping segments of size $n$ where $N_n = \lfloor N/n \rfloor$. For each segment, the root-mean-square function is calculated as:
\begin{equation}
    F_{\nu}(n) = \sqrt{\frac{1}{n} \sum_{i=1}^{n} \epsilon_{\nu}^2(i)}.
\end{equation}

The generalized fluctuation function of order $q$ is calculated as:
\begin{equation}
    F_q(n) = \left( \frac{1}{N_n} \sum_{v=1}^{N_n} \left[ F_{\nu}(n) \right]^q \right)^{1/q}, \quad q \neq 0.
\end{equation}

Where $q$ can take any real value except zero. For $q = 0$ the fluctuation function is given by:

\begin{equation}
    F_0(n) = \exp\left( \frac{1}{2N_n} \sum_{v=1}^{N_n} \ln \left[ F_{\nu}^2(n) \right] \right).
\end{equation}

This procedure is repeated for various segment sizes, to establish the relationship between the fluctuation function and segment size. If long-term correlations are present in the analyzed series, the fluctuation function increases with segment size according to the power law:
\begin{equation}
    F_q(n) \propto n^{h(q)},
\end{equation}

The scaling exponent $h(q)$ is referred to as the generalized Hurst exponent (for stationary time series it is identical to the classical Hurst exponent). It is obtained as the slope of linear regression of $\log F_q(n)$ versus $\log n$. The exponents describe the scaling behavior of subsets of the series with large fluctuations ($q > 0$) and subsets with small fluctuations ($q < 0$). For multifractal series, subsets with small and large fluctuations scale differently, and the exponent is a decreasing function of $q$. In contrast, for monofractal time series, it remains constant. Series with a larger range of exponent values exhibit a higher degree of multifractality \citep{kantelhardt2002multifractal}.

Multifractal properties of time series can also be characterized by the multifractal spectrum $f(\alpha)$, obtained through the Legendre transform:
\begin{equation}
    \alpha = \frac{d\tau(q)}{dq},
\end{equation}

\begin{equation}
    f(\alpha) = q\alpha - \tau(q),  
\end{equation}

A monofractal process is represented by a single point in the $f(\alpha)$ plane, while a multifractal process generates a single humped function \citep{kantelhardt2002multifractal}.

\begin{figure}
    \centering
    \includegraphics[width=\linewidth]{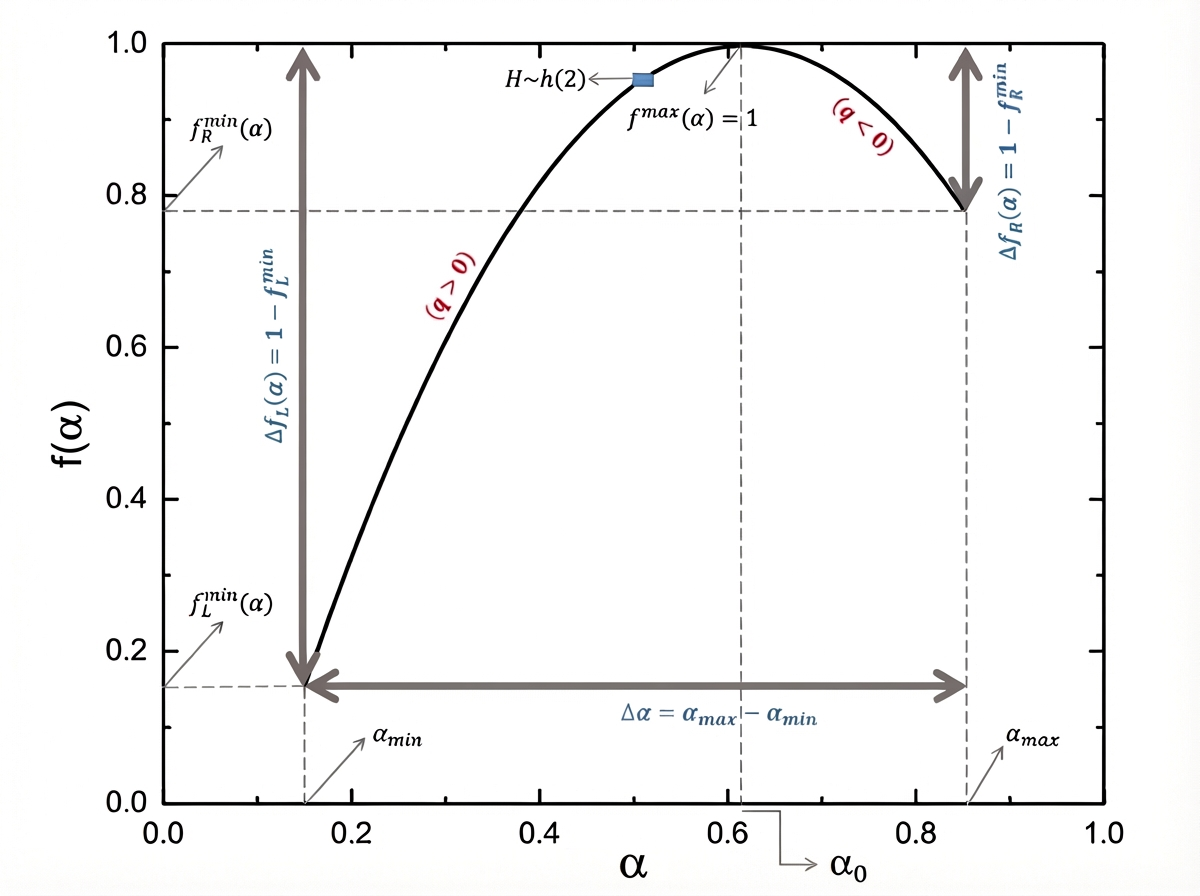}
    \caption{Multifractal singularity spectrum $f(\alpha)$ illustrating the position of the maximum at $\alpha_0$, the spectrum width $\Delta\alpha$, and the left and right contributions used to define the asymmetry of the spectrum.}
    \label{fig:PowerSpectrum}
\end{figure}

The complexity of an underlying stochastic process can be characterized by specific parameters of the $f(\alpha)$ spectrum: the position of the maximum $\alpha_0$, which corresponds to $f(\alpha_0) = f_{\max} = 1$; the spectrum width $\Delta \alpha = \alpha_{\max} - \alpha_{\min}$; and the asymmetry parameter $A = \Delta f_R / \Delta f_L$, where $\Delta f_R = 1 - f_R^{\min}$ and $\Delta f_L = 1 - f_L^{\min}$. These features are illustrated in Figure~\ref{fig:PowerSpectrum}. The parameter $A = 1$ indicates a symmetric spectrum, $A > 1$ corresponds to right-skewed shapes, and $A < 1$ represents left-skewed shapes. The parameter $h(q=2) = H$ describes the overall persistence or antipersistence of the underlying process. For persistent processes, larger values of $H$ indicate stronger persistence, whereas for antipersistent processes, smaller values of $H$ indicate stronger antipersistence. Values of $0 < H < 0.5$ and $0.5 < H < 1$ indicate long-range antipersistent and persistent correlations, respectively, while $H = 0.5$ corresponds to an uncorrelated time series \citep{feder1988multifractal, movahed2006multifractal}.

The width of the spectrum $\Delta\alpha$ represents the degree of multifractality in the process: larger $\Delta\alpha$ values signify a wider range of fractal exponents and stronger multifractality. The asymmetry parameter $A$ reflects the dominant fractal exponents. If the spectrum is right-skewed, the process is dominated by the scaling of small fluctuations (higher scaling exponents), and if the spectrum is left-skewed, the process is dominated by the scaling of large fluctuations (lower scaling exponents). Signals producing spectra with a wide range $\Delta\alpha$ (indicating a higher degree of multifractality) and a right-skewed shape (indicating the dominance of small fluctuations) may be considered more complex than those with opposite characteristics \citep{shimizu2002multifractal,drozdz2015detecting,telesca2004investigating}.

\subsection{ Fisher-Shannon Information Plane}
\label{sec:fisher_shannon}

The Fisher-Shannon analysis combines Shannon entropy and Fisher information measures to characterize the complexity of signals or probability distributions. This section introduces the key concepts and properties of these measures. The foundations of this analysis lie in two seminal works: \cite{shannon1948}, who introduced the concept of entropy in the context of information theory, and \cite{fisher1925}, who developed the theory of statistical estimation and introduced the Fisher information measure.

\subsubsection{Shannon Entropy Power and Fisher Information Measure}
Let \(X\) be a continuous random variable with probability density function (PDF) \(f(x)\). The differential entropy of \(X\) is defined as \cite{cover2006}:
\begin{equation}
    H_X = \mathbb{E}[-\log f(X)] = -\int f(x) \log f(x) \, dx.
    \label{eq:entropy}
\end{equation}
For a Gaussian random variable with variance \(\sigma^2\), \(H_X = \frac{1}{2} \log(2\pi e \sigma^2)\). The concept of entropy was first introduced~by \cite{shannon1948} in the context of information theory. The Shannon Entropy Power (SEP) is defined as \cite{dembo1991}:
\begin{equation}
    N_X = \frac{1}{2\pi e} e^{2H_X},
    \label{eq:sep}
\end{equation}
which, in the Gaussian case, simplifies to \(N_X = \sigma^2\). The SEP is a measure of uncertainty or spread of the distribution.

The Fisher Information Measure (FIM) is defined as \cite{vignat2003}:
\begin{equation}
    I_X = \mathbb{E}\left[\left(\frac{\partial}{\partial x} \log f(X)\right)^2\right] = \int \frac{\left[\frac{\partial}{\partial x} f(x)\right]^2}{f(x)} \, dx.
    \label{eq:fim}
\end{equation}
For a Gaussian distribution, \(I_X = 1/\sigma^2\). The FIM quantifies the order or narrowness of the distribution and has its roots in the work of \cite{fisher1925} on statistical estimation.

\subsubsection{Properties}
The SEP and FIM satisfy several properties \citep{dembo1991, rioul2011}:
\begin{itemize}
    \item Both are positive quantities.
    \item They scale as \(N_{aX} = a^2 N_X\) and \(I_{aX} = a^{-2} I_X\) for any real number \(a \neq 0\).
    \item They are invariant under additive constants.
\end{itemize}
The entropy power inequality and its dual, the Fisher information inequality, are given by \cite{dembo1991}:
\begin{align}
    N_{X+Y} &\geq N_X + N_Y, \label{eq:entropy_ineq} \\
    I_{X+Y}^{-1} &\geq I_X^{-1} + I_Y^{-1}, \label{eq:fisher_ineq}
\end{align}
for independent random variables \(X\) and \(Y\), with equality for Gaussian distributions. 

The product $Cx=NxIx$, called Fisher-Shannon Complexity (FSC)  \citep{angulo2008} is invariant under linear transformations and satisfies $CX \geq 1$, with equality for Gaussian distributions. It quantifies the sensitivity of $N_X$ to small additive perturbations.

\section{Results  and Discussion}\label{sec:results}
The data used in this study were extracted from \citet{edelson2019first} and correspond to observations of four quasars. Each of these objects was monitored across eight energy bands, ranging from the optical V filter to hard X-rays. The analyzed quasars exhibit redshifts in the range of $z = 0.0032$ to $z = 0.0341$ and show Eddington ratios varying between 1\% and 8\%.

In order to characterize the variability and complexity of the time series associated with each energy band, we employed two complementary approaches. The first is based on multifractal analysis using the MFDMA method, while the second relies on information theory, applying Fisher-Shannon analysis to quantify complexity.

We present below the application of these techniques to our sample, as previously discussed. The results are organized as follows: multifractal analysis is presented in Section~\ref{subsec:result_mfdma}, and the Fisher-Shannon complexity maps are discussed in Section~\ref{subsec:result_fs}.

\subsection{Multifractal Analysis}\label{subsec:result_mfdma}

\begin{figure*}
    \centering
    \includegraphics[width=13cm]{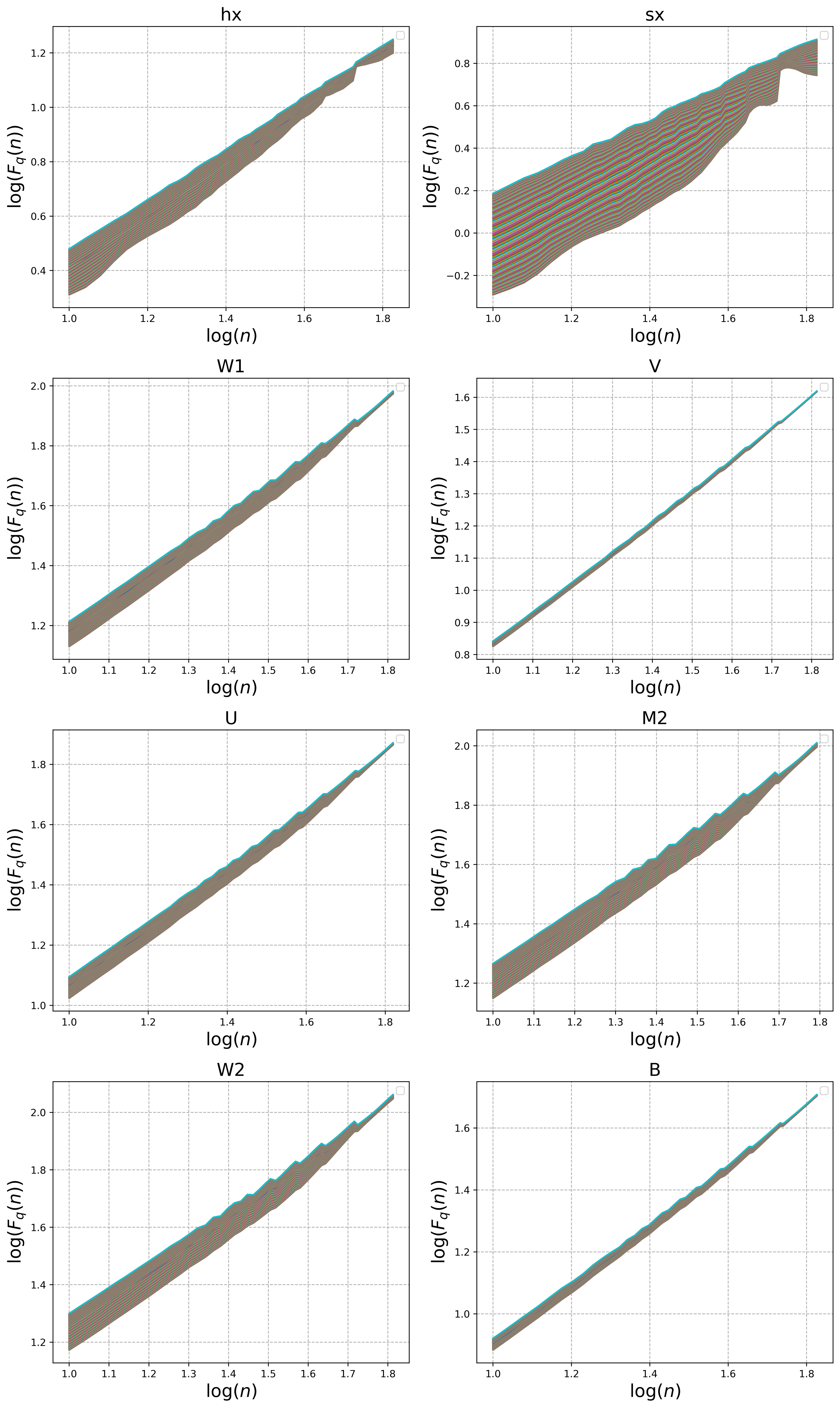}
      \caption{Fluctuation function $F_q(n)$ of the original light curve of NGC~5548 for all eight observational bands. The panels are arranged from left to right, top to bottom, in the following order: HX, SX, W2, M2, W1, U, B, and V. Each panel shows the scale dependence of fluctuations for different values of $q$, highlighting the variability characteristics across energy bands.}
    \label{fig:fq_plot}
\end{figure*}

At this point of the analysis, we begin by examining the fluctuation function $F_q(n)$ in order to understand how small- and large-scale fluctuations vary with the scale parameter $n$. Figure~\ref{fig:fq_plot} illustrates this relationship, showing that for different values of $q$, the scale dependence manifests differently across energy bands (for all AGNs see \ref{sec:appendix}). In particular, the soft X-ray (SX) band exhibits greater variation compared to the others, indicating a more complex and, in this case, multifractal profile. In contrast, the remaining bands display an essentially monofractal behavior. This  complexity in the SX band is consistent with findings in the literature, where the soft X-ray emission in Seyfert galaxies has been extensively associated with reverberation lags and coronal structure variability \citep{fabian2009broad, miller2010x, cackett2013soft, de2015tracing, wilkins2017revealing}. A similar multifractal behavior in the soft band was also reported for Mrk 421 by \citet{souza2025multifractal}.

The multifractal analysis comprises three main steps. First, we investigate how the generalized Hurst  exponent, \( h(q) \), varies with the moments \( q \), which modulate the amplitude across the studied scales. Next, we analyze how the scaling exponent, \( \tau(q) \), depends on \( q \). A linear relationship between \( \tau \) and \( q \) indicates monofractality, whereas a nonlinear dependence suggests multifractality. Finally, for each observed band, we compute the multifractal spectrum function \( f(\alpha) \) and determine the complexity parameters: the position of the maximum \( \alpha_{0}\), the spectrum width \(\Delta \alpha \), the asymmetry parameter \( A \), and  the Hurst exponent H. This analysis was extended to the entire sample, enabling a comparison of the energy bands across all quasars.

\begin{figure*}
    \centering
    \includegraphics[width=0.9\linewidth]{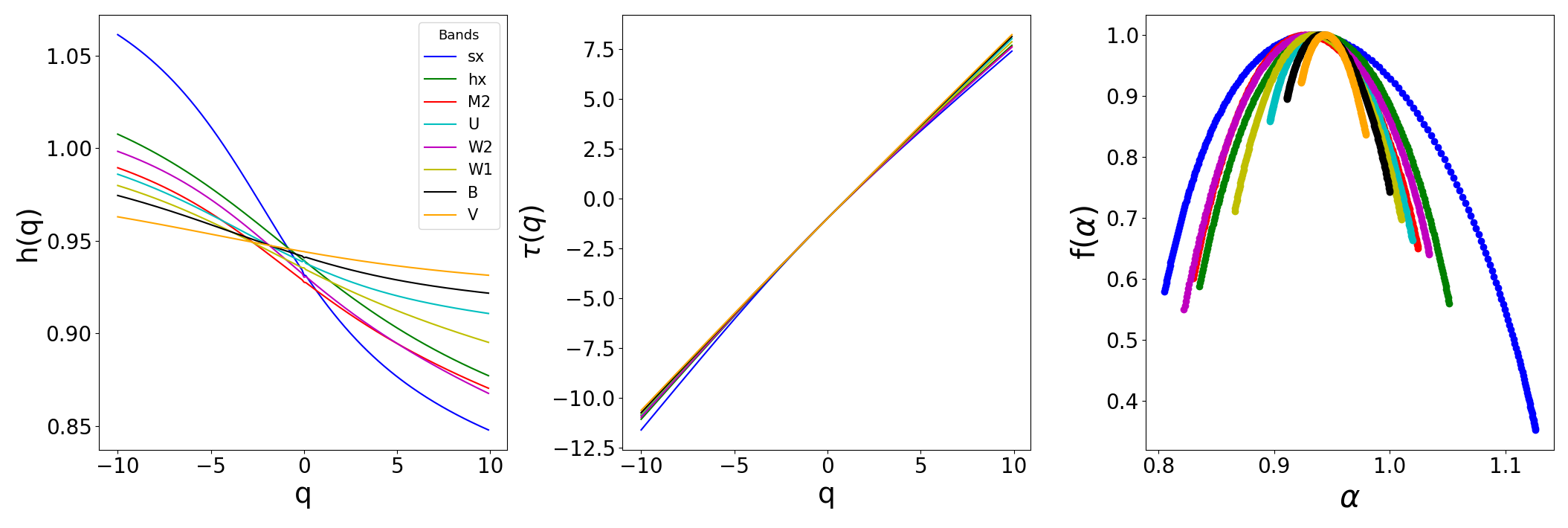}
        \includegraphics[width=0.9\linewidth]{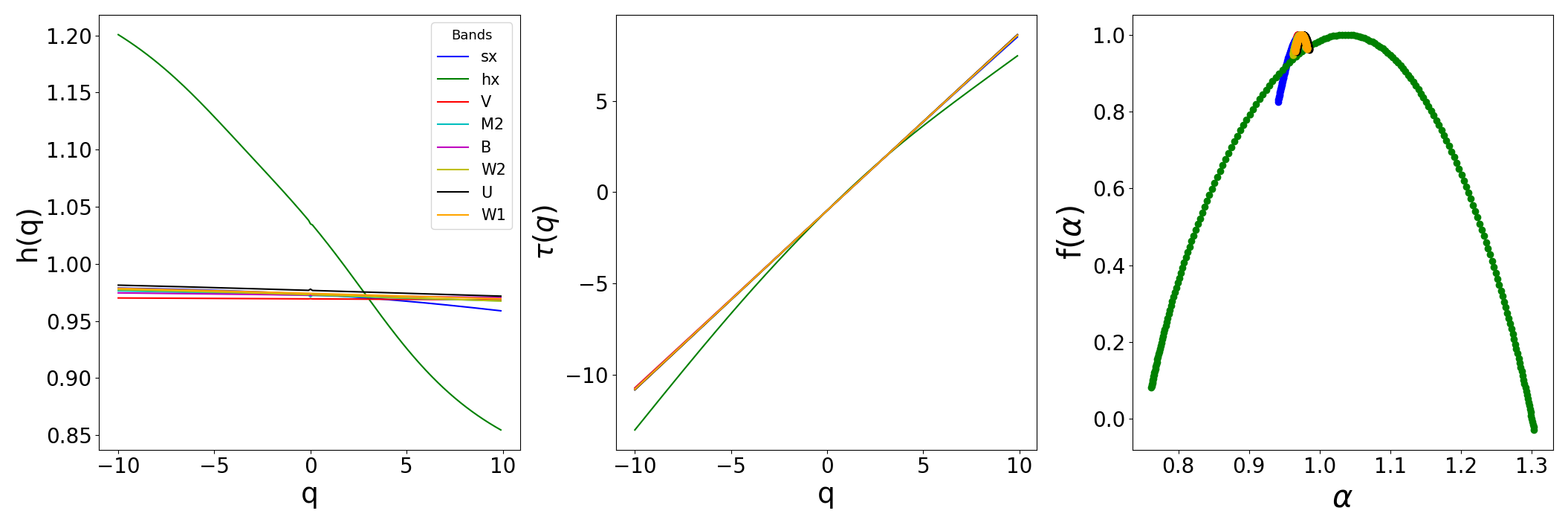}
                \includegraphics[width=0.9\linewidth]{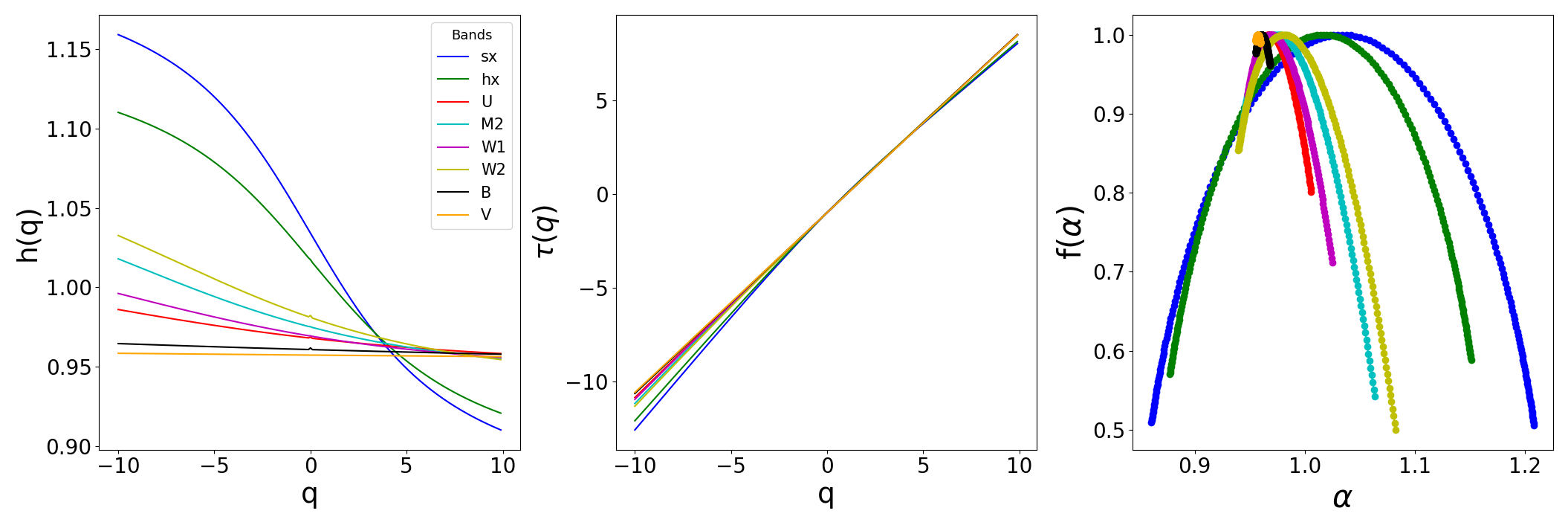}

        \includegraphics[width=0.9\linewidth]{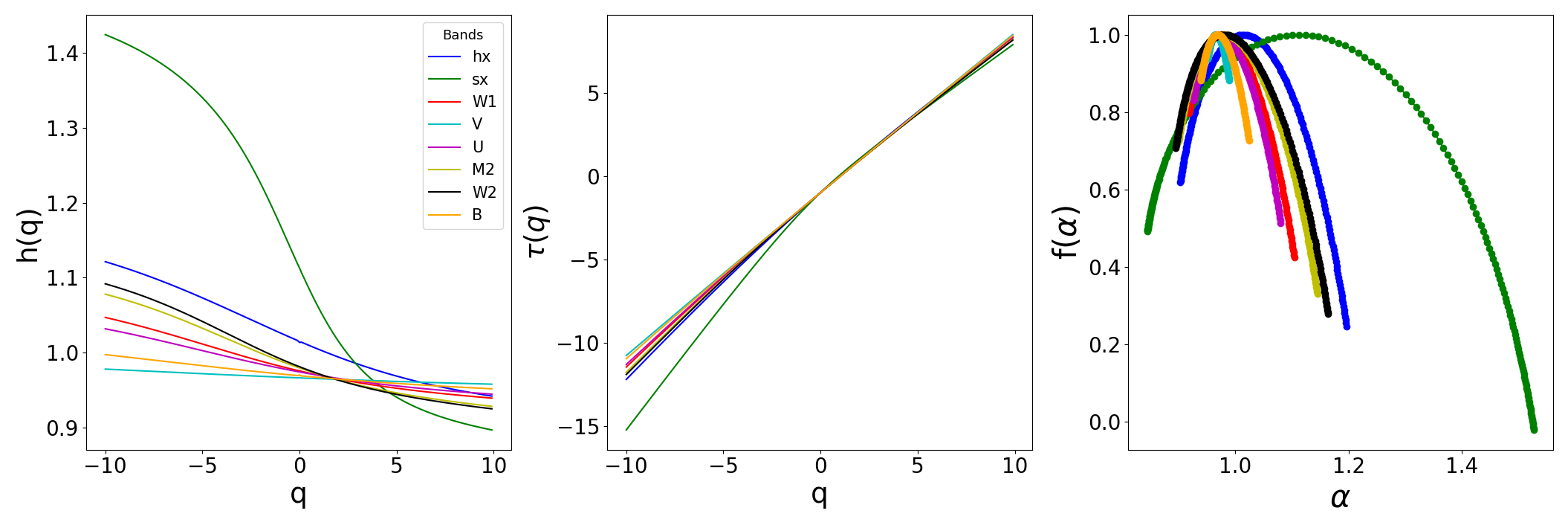}

\caption{Multifractal analysis using the MFDMA method for the four quasars in the sample, top to bottom: Mrk~509, NGC~4151, NGC~4593, and NGC~5548. Each figure consists of three panels, from left to right: the Hölder exponent $h(q)$ as a function of the moment $q$, the scaling exponent $\tau(q)$ versus $q$, and the multifractal spectrum $f(\alpha)$ as a function of the generalized Hölder exponent $\alpha$. These plots characterize the multifractal properties of the light curves for each object.}

    \label{fig:spectrum_mfdma}
\end{figure*}

Each quasar analyzed in this study exhibited multifractal characteristics, as shown in Figure~\ref{fig:spectrum_mfdma}. However, the multifractal features varied across the observational bands for each source. Therefore, it may be more appropriate to examine each quasar individually in order to better understand how multifractality manifests in different energy ranges. The values of multifractal parameters are presented in Table \ref{tab:main_results}. 

For the first quasar, Mrk 509, all energy bands except B and V exhibit multifractal behavior, with the effect being most prominent in the X-ray bands, SX and HX. Energy bands B and V display a very narrow spectrum $(\Delta \alpha \simeq 0)$ indicating monofractal behavior. The values of the asymmetry parameter A indicate that   small fluctuations are dominant $(A>1)$ for bands SX and U,  while for bands HX, W2, M2 and W1, $A \simeq 1$, suggesting that both small and large fluctuations contribute to the multifractality of the process. Additionally, the Hurst exponent $H$ is high and greater than 0.5 across all bands and all quasars in the sample, indicating strong persistence. This suggests that the fluctuation patterns remain consistent throughout the light curves across the entire dataset.

The second quasar, NGC~4151, stands out from the rest of the sample. Only the HX band exhibits clear multifractal properties, while the other bands display monofractal behavior, characterized by approximately constant $h(q)$, a linear $\tau(q)$ function, and a singular point in the $f(\alpha)$ spectrum (Figure~\ref{fig:spectrum_mfdma}), as well as low $\Delta \alpha$ values (Table~\ref{tab:main_results}). This may indicate the presence of additional or distinct physical processes driving the variability in the HX band. In contrast, the monofractal behavior observed in the other bands suggests that a single physical mechanism likely governs the variability across those energy ranges, rather than a combination of different processes.

The last two quasars, NGC~4593 and NGC~5548, show similar multifractal patterns. In NGC~4593, multifractality is more pronounced in the HX and SX bands compared to the optical ones. The W1,U, B and V bands display monofractal behavior, indicated by a very narrow multifractal spectrum $(\Delta \alpha \simeq 0)$. Notably, the multifractal spectrum in the optical bands W2 and M2 reveals right asymmetry $(A > 1)$, indicating that the scaling of small fluctuations dominates the multifractality of the process. For NGC~5548, multifractality is also more pronounced in the HX and SX bands, while B and V bands display monofractal behavior $(\Delta \alpha \simeq 0)$. A strong right $(A>1)$ asymmetry is also observed across the entire set of bands, indicating the dominance of small fluctuations. The SX band, in particular, exhibits the most prominent multifractality (highest value of spectrum width $\Delta \alpha=0.681$)  among all bands of quasars analyzed in this study.

Physically, perturbations, such as intrinsic instabilities or variations in the accretion rate, originating in the innermost regions of the AGN can propagate outwards to more extended regions, maintaining both temporal and spatial coherence. This evidence reinforces the notion that AGN variability does not stem from stochastic behavior but is, rather, a manifestation of nonlinear and non-stationary mechanisms exhibiting significant temporal correlations \citep{edelson2015space,edelson2019first}. This includes crucial processes like radiation reprocessing and reverberation between distinct emission regions, which serve as the causal links for the observed long-term variability. Notably, soft X-ray bands show stronger multifractality than optical bands, reflecting more complex temporal structures in higher-energy regimes. The multifractal nature of these energy bands is consistent with the influence of thermally unstable or reprocessed emission from more extended regions of the accretion disk, or from X-ray heated winds, which induce multi-scale fluctuations and intermittency\citep{Longo1996,vaughan2003characterizing,belete2019novel}.

\begin{figure}

    \centering
    \includegraphics[width=\linewidth]{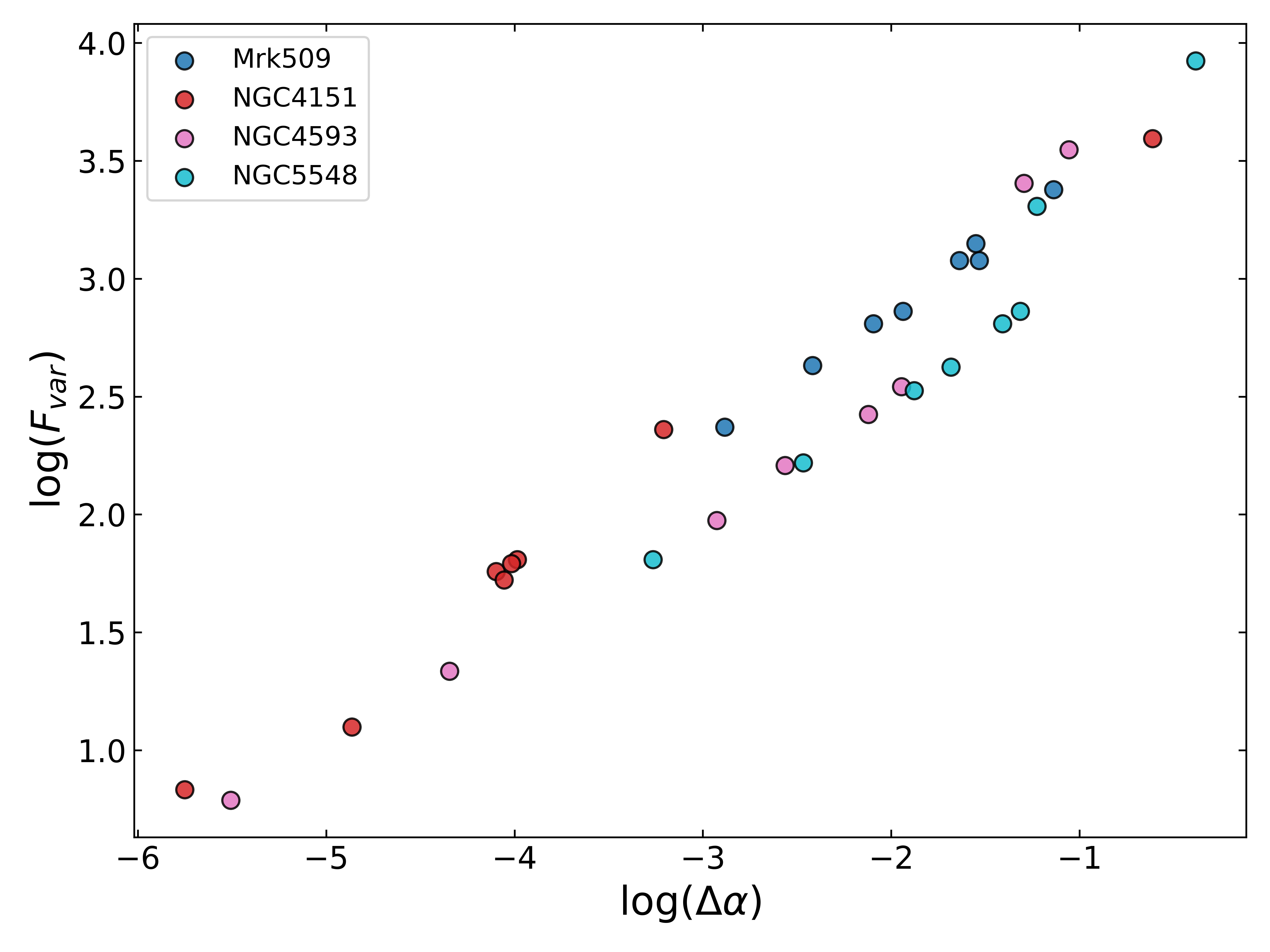}
    \caption{Scatter plot of fractional variability as a function of $\Delta \alpha$ for different AGNs. Each color represents a distinct AGN. The plot reveals a strong positive correlation between the two quantities, suggesting that larger spectral widths are associated with greater variability amplitudes.}

    \label{fig:fvar_vs_w}
\end{figure}

A metric introduced by \citet{vaughan2003characterizing}, the fractional variability $F_{\mathrm{var}}$, quantifies the excess variance of a time series and is commonly used to characterize quasar variability, which is inherently stochastic and thus difficult to describe deterministically. In this work, we present a measure that is clearly related to $F_{\mathrm{var}}$, namely the degree of multifractality ($ \Delta \alpha$). This quantity offers a richer interpretation, as it is intrinsically connected to the amplitude of variability captured by $F_{\mathrm{var}}$.
In \citet{souza2025multifractal}, an energy-dependent behavior of $\Delta \alpha$ is reported for the blazar Mrk 421, suggesting that this metric may behave similarly to the $F_{\mathrm{var}}$. The monotonic relationship between the two metrics (presented in Table~\ref{tab:main_results}), as illustrated in Figure~\ref{fig:fvar_vs_w}, highlights an intrinsic connection between multifractal characteristics and the amplitude of variability.

We investigate the relationship between the multifractal width $\Delta \alpha$ and the fractional variability $F_{\mathrm{var}}$ by modeling their dependence through a power-law relation of the form $\Delta \alpha \propto F_{\mathrm{var}}^{\beta}$ (Figure~\ref{fig:fvar_vs_w}). The scaling exponent $\beta$ is estimated via linear regression in logarithmic space, such that $\log \Delta \alpha = \beta \log F_{\mathrm{var}} + C$, where $C$ is a normalization constant. The fit yields $\beta = 1.69 \pm 0.08$, with a high coefficient of determination ($R^{2} = 0.94$), indicating that the power-law model provides an excellent description of the data.

The value of the scaling exponent $\beta = 1.69$ indicates that $\Delta\alpha$ responds in a \emph{superlinear} manner to variations in $F_{\mathrm{var}}$. In a superlinear power-law relation, relative changes in the independent variable lead to amplified relative changes in the dependent one, reflecting enhanced sensitivity. Such behavior is commonly interpreted as a signature of intensified interactions or nonlinear responses within complex systems and has been widely discussed in the context of scaling phenomena \citep{Yakubo2014UrbanScaling,Shi2022UrbanResilienceScaling}. Consequently, within the observed regime, $\Delta\alpha$ can be regarded as a particularly sensitive indicator of variability-related dynamical complexity.

\begin{table}
    \centering
    \caption{Main parameters for different observational bands of the analyzed QSOs. $\Delta \alpha$ represents the degree of multifractality, $A$ is the asymmetry of the multifractal spectrum, $\alpha_0$ denotes the position of the spectrum's maximum, and $H$ is the local Hurst exponent. $F_{\mathrm{var}}$ (\%) corresponds to the fractional variability, with values taken from \citet{edelson2019first}.}

    \begin{tabular}{lcccccc}
        \hline
        QSO & Band & $\Delta \alpha$ & A & $\alpha_0$ & H & $F_{\mathrm{var}}$ (\%)\\
        \hline
        Mrk~509 & HX & 0.216 & 1.084 & 0.939 & 0.924 & 21.70 \\
        Mrk~509 & SX & 0.321 & 1.519 & 0.933 & 0.906 & 29.30 \\
        Mrk~509 & W2 & 0.212 & 0.939 & 0.931 & 0.915 & 23.30 \\
        Mrk~509 & M2 & 0.194 & 0.982 & 0.928 & 0.913 & 21.70 \\
        Mrk~509 & W1 & 0.144 & 1.089 & 0.935 & 0.925 & 17.50 \\
        Mrk~509 & U  & 0.123 & 1.944 & 0.938 & 0.930 & 16.60 \\
        Mrk~509 & B  & 0.089 & 1.937 & 0.941 & 0.936 & 13.90 \\
        Mrk~509 & V  & 0.056 & 1.707 & 0.944 & 0.941 & 10.70 \\
        \hline 
        NGC~4151 & HX & 0.542 & 0.975 & 1.036 & 0.993 & 36.40 \\
        NGC~4151 & SX & 0.040 & 0.264 & 0.973 & 0.971 & 10.60 \\
        NGC~4151 & W2 & 0.019 & 0.719 & 0.973 & 0.972 & 6.10 \\
        NGC~4151 & M2 & 0.017 & 0.765 & 0.972 & 0.971 & 5.80 \\
        NGC~4151 & W1 & 0.017 & 0.896 & 0.974 & 0.973 & 5.60 \\
        NGC~4151 & U  & 0.018 & 0.885 & 0.977 & 0.976 & 6.00 \\
        NGC~4151 & B  & 0.008 & 0.913 & 0.973 & 0.972 & 3.00 \\
        NGC~4151 & V  & 0.003 & 0.895 & 0.969 & 0.969 & 2.30 \\
        \hline 
        NGC~4593 & HX & 0.274 & 0.961 & 1.017 & 0.989 & 30.10 \\
        NGC~4593 & SX & 0.348 & 1.006 & 1.034 & 0.994 & 34.70 \\
        NGC~4593 & W2 & 0.143 & 2.456 & 0.981 & 0.973 & 12.70 \\
        NGC~4593 & M2 & 0.120 & 2.836 & 0.975 & 0.969 & 11.30 \\
        NGC~4593 & W1 & 0.077 & 2.602 & 0.969 & 0.966 & 9.10 \\        
        NGC~4593 & U  & 0.054 & 2.394 & 0.968 & 0.965 & 7.20 \\
        NGC~4593 & B  & 0.013 & 1.454 & 0.961 & 0.960 & 3.80 \\
        NGC~4593 & V  & 0.004 & 1.221 & 0.957 & 0.957 & 2.20 \\
        \hline 
        NGC~5548 & HX & 0.293 & 1.630 & 1.015 & 0.994 & 27.30 \\
        NGC~5548 & SX & 0.681 & 1.548 & 1.113 & 0.015 & 50.60 \\
        NGC~5548 & W2 & 0.269 & 2.135 & 0.981 & 0.963 & 17.50 \\
        NGC~5548 & M2 & 0.244 & 2.092 & 0.979 & 0.963 & 16.60 \\
        NGC~5548 & W1 & 0.186 & 2.276 & 0.975 & 0.965 & 13.80 \\
        NGC~5548 & U  & 0.153 & 2.267 & 0.974 & 0.965 & 12.50 \\
        NGC~5548 & B  & 0.085 & 1.890 & 0.969 & 0.964 & 9.20 \\
        NGC~5548 & V  & 0.038 & 1.598 & 0.966 & 0.964 & 6.10 \\
        \hline
    \end{tabular}
    \label{tab:main_results}
\end{table}

\subsection{Fisher-Shannon Analysis} \label{subsec:result_fs}
In this stage, we compute the Shannon entropy power ($N_X$) and the Fisher information ($I_X$) for each of the time series. Using these quantities, we construct the complexity plane based on their joint behavior \citep{li2020spectral, aguiar2023quantifying}.

\begin{figure}
    \centering
        \includegraphics[width=\linewidth]{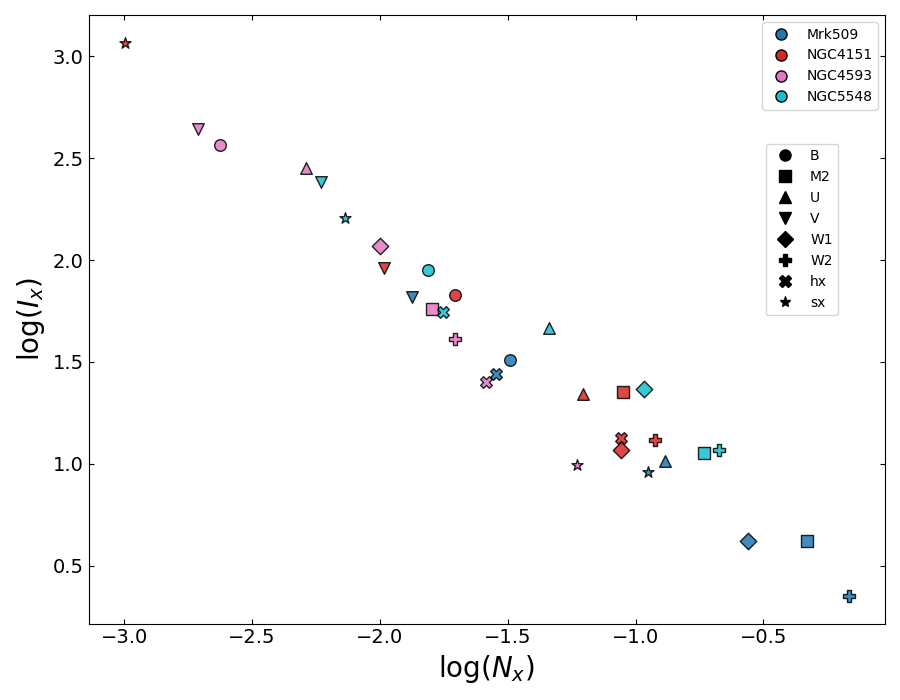}
\caption{Fisher–Shannon complexity plane defined by $\log(N_x)$ and $\log(I_x)$ for the four Seyfert galaxies analyzed, where colors denote the sources and symbols represent the observational bands.}

\label{fig:plane_complex}
\end{figure}

The Fisher-Shannon (FS) information plane, as shown in Figure~\ref{fig:plane_complex}, offers a two-dimensional framework for analyzing the dynamical complexity of time series by combining the Fisher information measure ($I_X$) with the Shannon entropy power ($N_X$) \citep{pierini2015discriminating}. In this representation, each point corresponds to a quasar light curve observed in a specific energy band. A distinct anti-correlation is observed: as $N_X$ increases - indicating higher uncertainty or disorder - the value of $I_X$ tends to decrease, reflecting a reduction in structural organization. This distribution highlights intrinsic differences in the variability behavior across spectral bands and sources.

Notably, the upper-left region of the FS plane, characterized by low $\log(N_X)$ and high $\log(I_X)$ values, is populated by soft X-ray (SX) observations such as those from NGC~4151 and optical bands like V and B from NGC~4593. These signals exhibit highly ordered temporal structures, suggesting emission from compact and coherent regions near the black hole corona.

Conversely, the lower-right region - marked by high $\log(N_X)$ and low $\log(I_X)$ - includes observations from Mrk~509 (W2, M2, W1, U, SX), NGC~5548 (M2, W2, W1), and NGC~4151 (W2), especially in the UV bands. These bands display more stochastic dynamics, likely due to the influence of thermally unstable or reprocessed emission from extended regions of the accretion disk. Therefore, the FS plane effectively discriminates between emission regimes with distinct levels of complexity and temporal organization.

\subsection{Can Complexity Measures Reveal Insights into Quasar Variability?} \label{subsec:result_complex_variability}

Complexity measures, particularly the multifractal width $\Delta \alpha$, provide valuable insight into quasar variability. $\Delta \alpha$ are not an isolated metric; the values of $\alpha$ encapsulate information about the nature of fluctuations at both small and large temporal scales, as well as the persistence or anti-persistence of these fluctuations. The observed data reveal a consistent positive relationship between $\Delta \alpha$ and the fractional variability $F_{\mathrm{var}}$. For instance, bands with higher $\Delta \alpha$, such as 0.681 and 0.542, correspond to higher $F_{\mathrm{var}}$ values (50.6\% and 36.4\%, respectively), indicating a monotonic association between the amplitude of variability and the degree of multifractality. This suggests that quasars exhibiting larger variability amplitudes tend to have more complex, scale-dependent fluctuation structures.

Moreover, the Fisher-Shannon complexity analysis supports this interpretation. The distribution of information quantifiers across energy bands and quasars predominantly falls between regions of low entropy with high Fisher information (low complexity) and the opposite regime, suggesting that the light curves exhibit substantial structure and information content. This implies that the signals are not entirely stochastic in all bands; for instance, bands with high $N_X$ and low $I_X$, such as Mrk 509 (W2, M2, W1, U, SX), NGC 5548 (M2, W2, W1), and NGC 4151 (W2), tend to display more stochastic variability associated with extended or thermally unstable emission regions. In contrast, bands with low $N_X$ and high $I_X$, like the soft X-ray (SX) band of NGC 4151 and the optical bands V and B of NGC 4593, exhibit more ordered, coherent temporal structures, reflecting emission dominated by compact regions near the black hole. Therefore, the FS plane provides a valuable tool for differentiating variability regimes, enabling a more detailed understanding of the physical processes driving quasar light curve dynamics.

Thus, complexity measures do not merely quantify irregularity; they also correlate with physical characteristics of variability, offering a more nuanced and interpretable framework than traditional variance-based approaches. In this regard, we acknowledge the concerns raised by \citet{assis2024multifractality}, that multifractality may be influenced by observational noise and missing data (i.e., gaps in light curves).  Nevertheless, our results demonstrate that multifractality and its derived metrics maintain a robust association with the well-established variability indicator \(F_{\mathrm{var}}\), thereby underscoring their physical relevance and mitigating concerns that these features arise solely from data quality issues.

\section{Conclusions}\label{sec:conclusion}

This study presents a comprehensive analysis of quasar variability using multifractal detrended moving average analysis and the Fisher–Shannon plane applied to multiwavelength light curves of Mrk~509, NGC~5548, NGC~4151, and NGC~4593. All sources exhibit long-term persistent correlations, with Hurst exponents consistently above 0.5. 
The observed long-term correlations indicate coherent signal propagation across the accretion disk, potentially associated with reverberation processes. These results support the view that AGN variability arises from mechanisms characterized by strong temporal correlations. 
In parallel, the superlinear exponent ($\beta = 1.69$) suggests that $\Delta\alpha$ may encode variability-related information similar to that captured by $F_{\mathrm{var}}$, potentially with enhanced sensitivity; however, larger datasets are required to robustly confirm this interpretation.

The optical B and V bands tend toward near-monofractal behavior, reflecting relatively stable variability. This reduced complexity can be interpreted as a result of more stable reprocessing processes in more extended, lower-energy regions of the accretion disk or the broad-line region (BLR), where fluctuations are characterized by lower intensity and higher predictability. For Mrk~509, NGC~4593, and NGC~5548, the multifractal width $\Delta \alpha$ is larger in soft X-rays than in hard X-rays, and in UV W2 than W1 bands, suggesting that emission from more compact or energetic regions displays richer scaling properties. This implies that even small perturbations can propagate through the system and generate variability across a broad range of temporal scales.

 These bands are characterized by stronger multifractality, conversely, the soft X-ray band of NGC 4151 and optical bands of NGC 4593 lie in regions of lower entropy and higher information, indicating more coherent emission likely originating from compact structures near the corona or BLR. These bands exhibiting lower degree of multifractality (near-monofractal behavior)  This consistency indicates a more coherent and ordered emission, which can be associated with compact and well-defined structures located in the vicinity of the black hole corona or the broad-line region. Thus, the FS plane not only quantifies variability irregularity but also correlates directly with its intrinsic physical characteristics, offering a more nuanced interpretative framework compared to approaches based solely on traditional variance, and validating inferences regarding the emission regions of AGNs.
Fisher-Shannon analysis complements these findings by mapping each light curve on the entropy -- information plane. An inverse relationship is evident: sources with high entropy (greater disorder) show low Fisher information (less structural detail), and vice versa. UV and soft X-ray bands of Mrk 509 and NGC 5548 occupy regions of high stochasticity, consistent with reprocessed or unstable emission likely originating in extended regions of the accretion disk or X-ray-heated winds. These bands are characterized by stronger multifractality, conversely, the soft X-ray band of NGC 4151 and optical bands of NGC 4593 lie in regions of lower entropy and higher information, indicating more coherent emission likely originating from compact structures near the corona or BLR. These bands exhibiting lower degree of multifractality (near-monofractal behavior). This consistency indicates a more coherent and ordered emission, which can be associated with compact and well-defined structures located in the vicinity of the black hole corona or the broad-line region. Thus, the FS plane not only quantifies variability irregularity but also correlates directly with its intrinsic physical characteristics, offering a more nuanced interpretative framework compared to approaches based solely on traditional variance, and validating inferences regarding the emission regions of AGNs.

In summary, complexity measures such as $\Delta \alpha$, Fisher information, and Shannon entropy provide physical explanations into AGN variability mechanisms. These metrics correlate well with classical variability indicators like $F_{\mathrm{var}}$, enabling discrimination between stochastic and structured emission processes. Despite potential sensitivity to noise and observational gaps \citep{assis2024multifractality}, consistent patterns observed across multiple bands and sources support the utility of these measures as diagnostic tools. Overall,this work enhances our understanding of how diverse physical processes shape quasar light curves across the electromagnetic spectrum.

\section*{Acknowledgements}
We acknowledge the University of Pernambuco (UPE) for their support and resources. We also acknowledge the Graduate Program in Applied Physics (PPGFA) at the Federal Rural University of Pernambuco (UFRPE) for their support and resources throughout this research. The authors acknowledge support of Brazilian agency CNPq through the following research grants: No 309499/2022-4. EH has received funding from the European Union's Horizon Europe research and innovation program under grant agreement No. 101188037 (AtLAST2). This research has made use of the VizieR catalogue access tool, CDS, Strasbourg, France (DOI : 10.26093/cds/vizier). The original description of the VizieR service was published in 2000, A\&AS 143, 23.

%%%%%%%%%%%%%%%%%%%%%%%%%%%%%%%%%%%%%%%%%%%%%%%%%%
\section*{Data Availability}

The data used in this study are available in the VizieR catalogue at \url{https://doi.org/10.26093/cds/vizier.18700123}. Additionally, the codes for the implementation of the MFDMA and Fisher–Shannon methods are available at the following repositories: \url{https://github.com/rhimonsouza/mfdma} and \url{https://github.com/fishinfo/FiShPy}, respectively.



\bibliographystyle{mnras}
\bibliography{example} % if your bibtex file is called example.bib

%\begin{thebibliography}{99}
%\end{thebibliography}

%%%%%%%%%%%%%%%%%%%%%%%%%%%%%%%%%%%%%%%%%%%%%%%%%%

%%%%%%%%%%%%%%%%% APPENDICES %%%%%%%%%%%%%%%%%%%%%

\appendix
\onecolumn

% ---- APPENDIX A ----
\clearpage
\section{Time Series for each AGN}
{\centering 
\noindent\includegraphics[width=\textwidth,height=0.80\textheight,keepaspectratio]{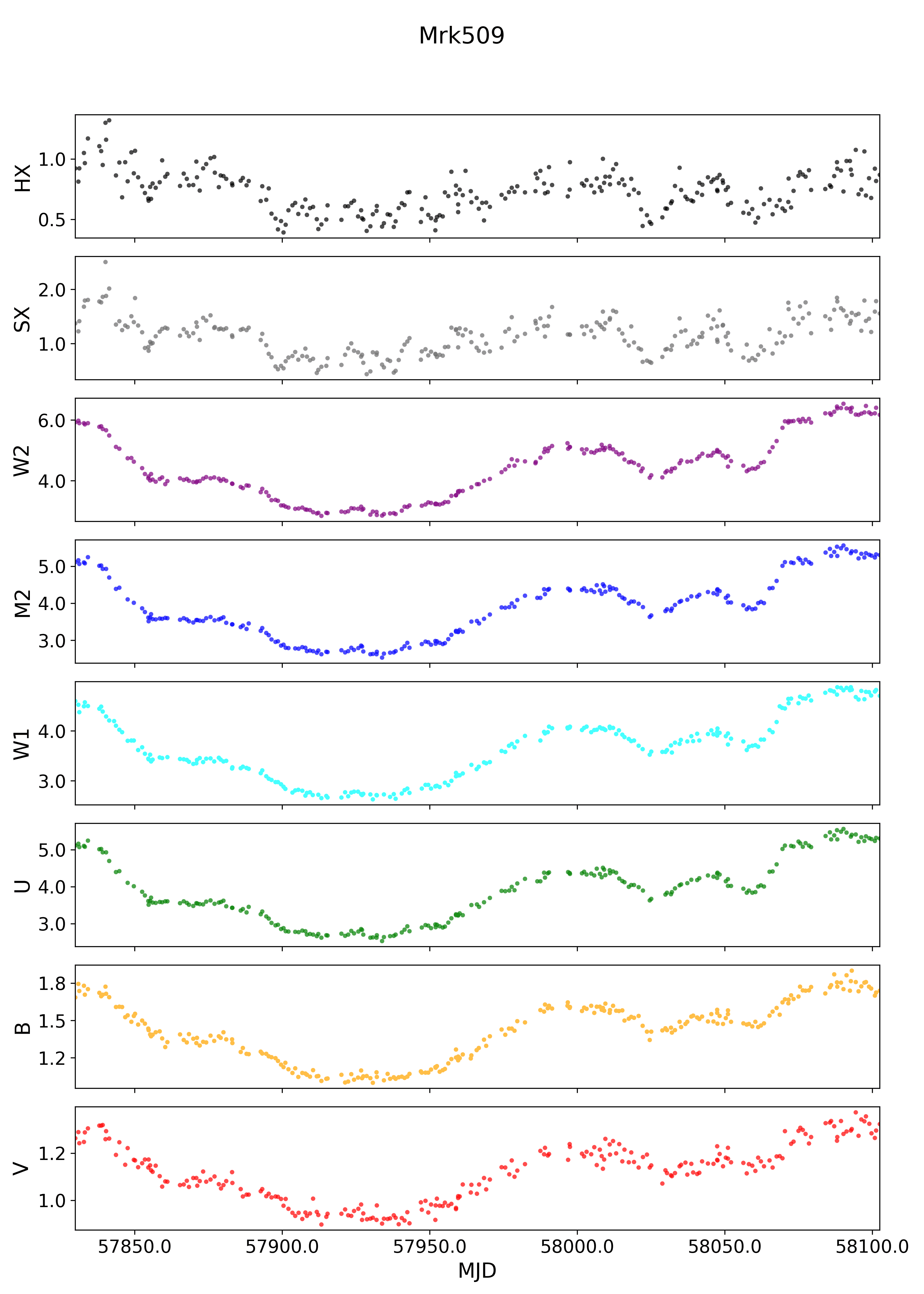}
\captionsetup{hypcap=false}\captionof{figure}{The original light curve of Mrk 509 for all eight observational bands. Light curves of the analyzed AGN, ordered by wavelength from top to bottom. The top two panels correspond to XRT observations in the HX (1.5--10 keV) and SX (0.3--1.5 keV) bands, while the lower six panels show the UVOT bands (W2, M2, W1, U, B, and V). Each panel displays the flux as a function of MJD, sharing a common time axis but with independent flux scales. X-ray fluxes are given in units of counts s$^{-1}$, and UVOT fluxes are expressed in units of $10^{-14}$\,erg\,cm$^{-2}$\,s$^{-1}$\,\AA$^{-1}$.}
\label{fig:lc_plot_Mrk509}

\clearpage

\noindent\includegraphics[width=\textwidth,height=0.80\textheight,keepaspectratio]{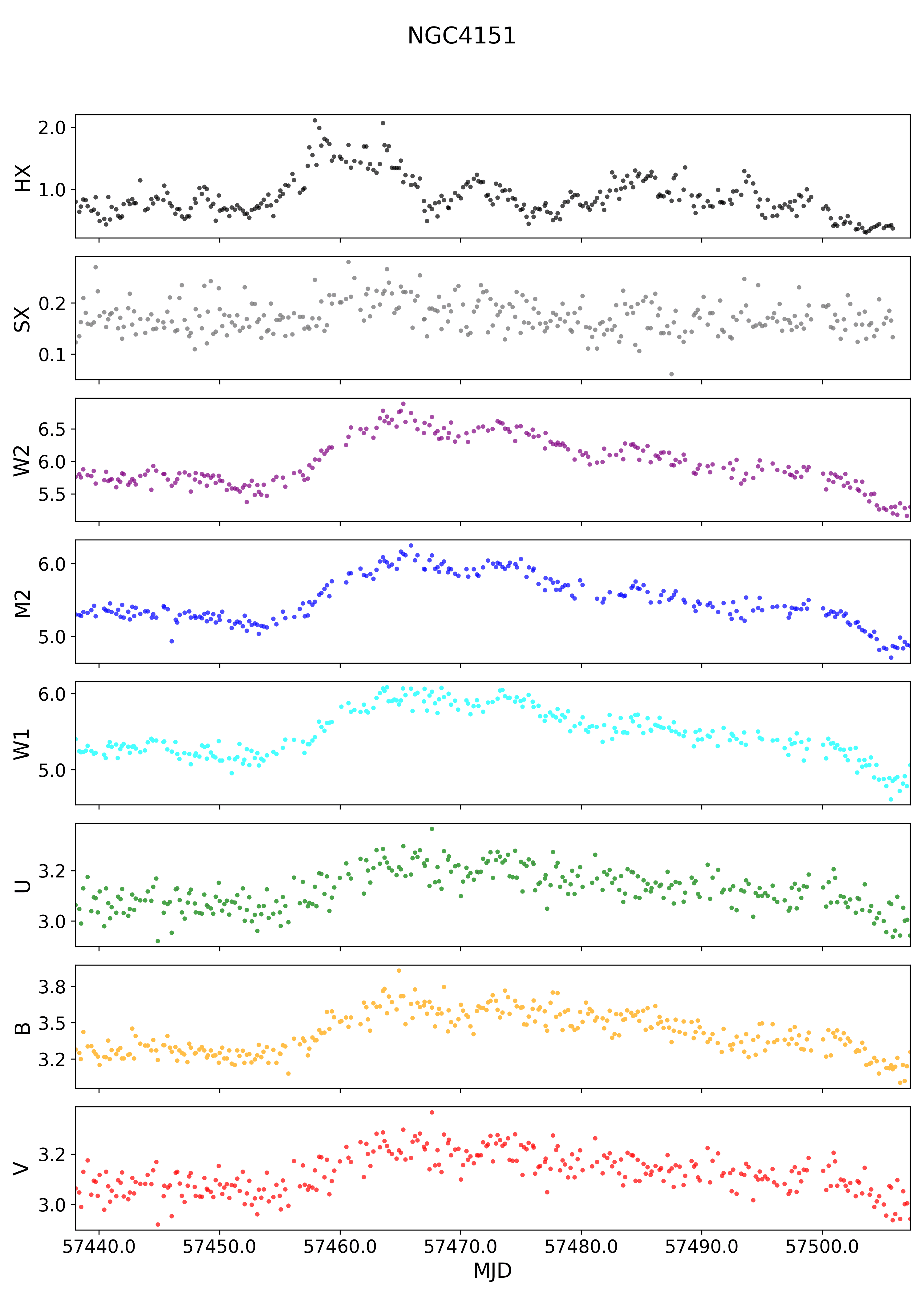}
\captionsetup{hypcap=false}\captionof{figure}{The original light curve of NGC 4151 for all eight observational bands. Light curves of the analyzed AGN, ordered by wavelength from top to bottom. The top two panels correspond to XRT observations in the HX (1.5--10 keV) and SX (0.3--1.5 keV) bands, while the lower six panels show the UVOT bands (W2, M2, W1, U, B, and V). Each panel displays the flux as a function of MJD, sharing a common time axis but with independent flux scales. X-ray fluxes are given in units of counts s$^{-1}$, and UVOT fluxes are expressed in units of $10^{-14}$\,erg\,cm$^{-2}$\,s$^{-1}$\,\AA$^{-1}$.}
\label{fig:lc_plot_NGC4151}

\clearpage

\noindent\includegraphics[width=\textwidth,height=0.80\textheight,keepaspectratio]{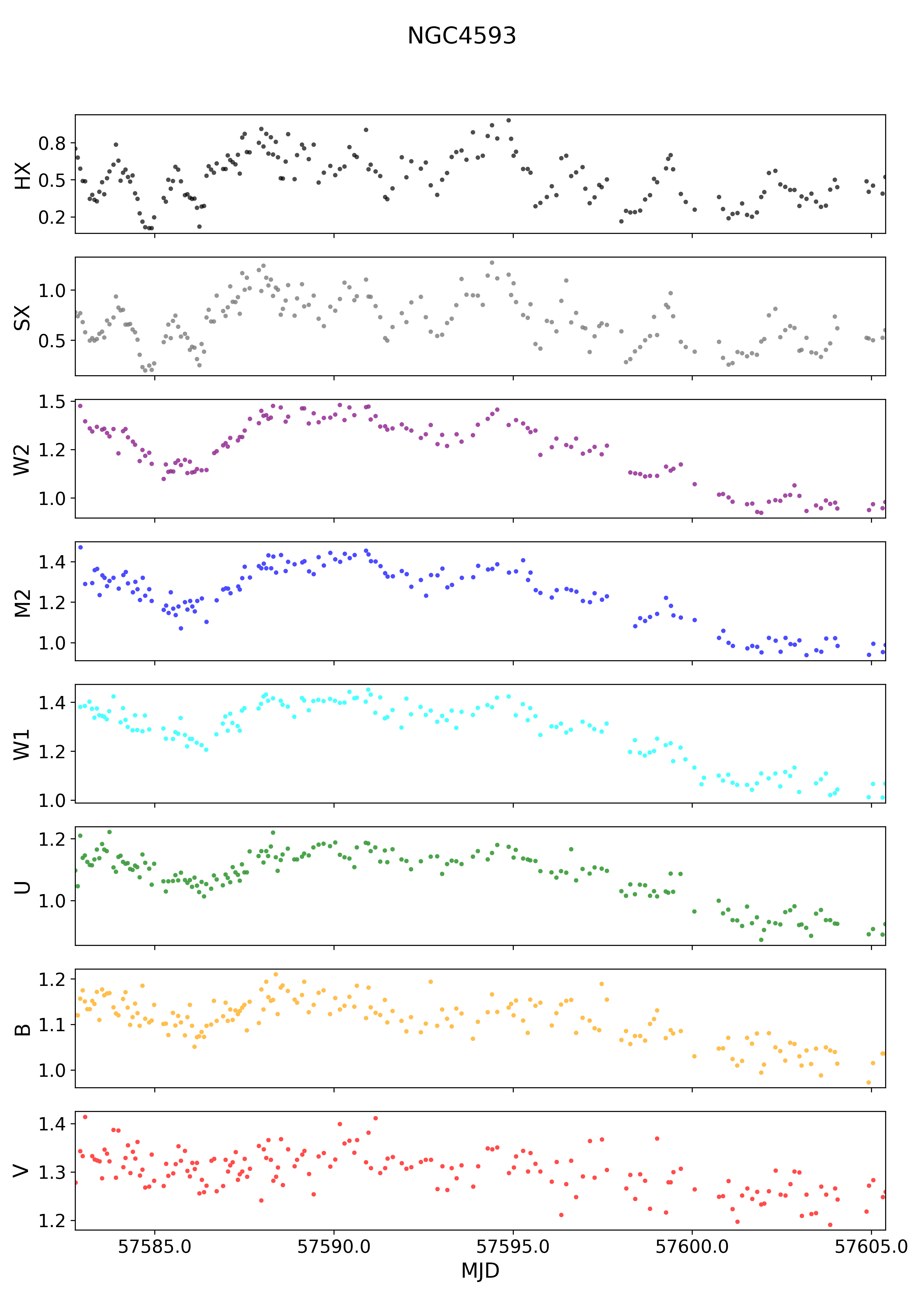}
\captionsetup{hypcap=false}\captionof{figure}{The original light curve of NGC 4593 for all eight observational bands. Light curves of the analyzed AGN, ordered by wavelength from top to bottom. The top two panels correspond to XRT observations in the HX (1.5--10 keV) and SX (0.3--1.5 keV) bands, while the lower six panels show the UVOT bands (W2, M2, W1, U, B, and V). Each panel displays the flux as a function of MJD, sharing a common time axis but with independent flux scales. X-ray fluxes are given in units of counts s$^{-1}$, and UVOT fluxes are expressed in units of $10^{-14}$\,erg\,cm$^{-2}$\,s$^{-1}$\,\AA$^{-1}$.}
\label{fig:lc_plot_NGC4593}

\clearpage

\noindent\includegraphics[width=\textwidth,height=0.80\textheight,keepaspectratio]{figures/AGN_Light_Curves/NGC5548_scatter_lightcurves.png}
\captionsetup{hypcap=false}\captionof{figure}{The original light curve of NGC~5548 for all eight observational bands. Light curves of the analyzed AGN, ordered by wavelength from top to bottom. The top two panels correspond to XRT observations in the HX (1.5--10 keV) and SX (0.3--1.5 keV) bands, while the lower six panels show the UVOT bands (W2, M2, W1, U, B, and V). Each panel displays the flux as a function of MJD, sharing a common time axis but with independent flux scales. X-ray fluxes are given in units of counts s$^{-1}$, and UVOT fluxes are expressed in units of $10^{-14}$\,erg\,cm$^{-2}$\,s$^{-1}$\,\AA$^{-1}$.}
\label{fig:lc_plot_5548}
\par}

% ---- APPENDIX B ----
\clearpage
\section{Fluctuation Function for each AGN}
\label{sec:appendix}
{\centering

\noindent\includegraphics[width=\textwidth,height=0.80\textheight,keepaspectratio]{figures/MFDMA/NGC5548/Fq_vs_n_2x2NGC5548.png}
\captionsetup{hypcap=false}\captionof{figure}{Fluctuation function $F_q(n)$ of the original light curve of NGC~5548 for all eight observational bands.}
\label{fig:fq_plot_1}

\clearpage

\noindent\includegraphics[width=\textwidth,height=0.80\textheight,keepaspectratio]{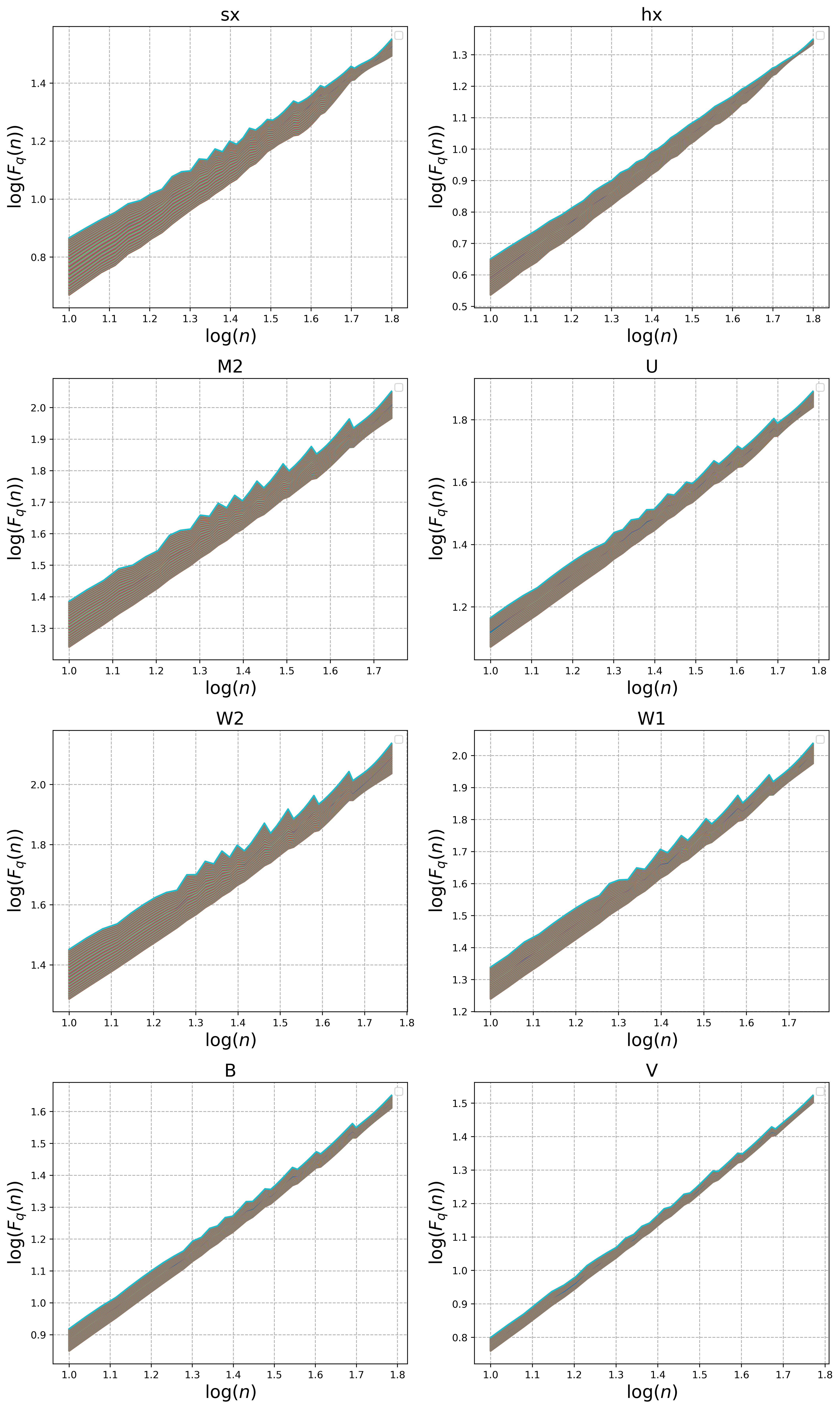}
\captionsetup{hypcap=false}\captionof{figure}{Fluctuation function $F_q(n)$ of the original light curve of Mrk~509 for all eight observational bands.}
\label{fig:fq_plot_2}

\clearpage

\noindent\includegraphics[width=\textwidth,height=0.80\textheight,keepaspectratio]{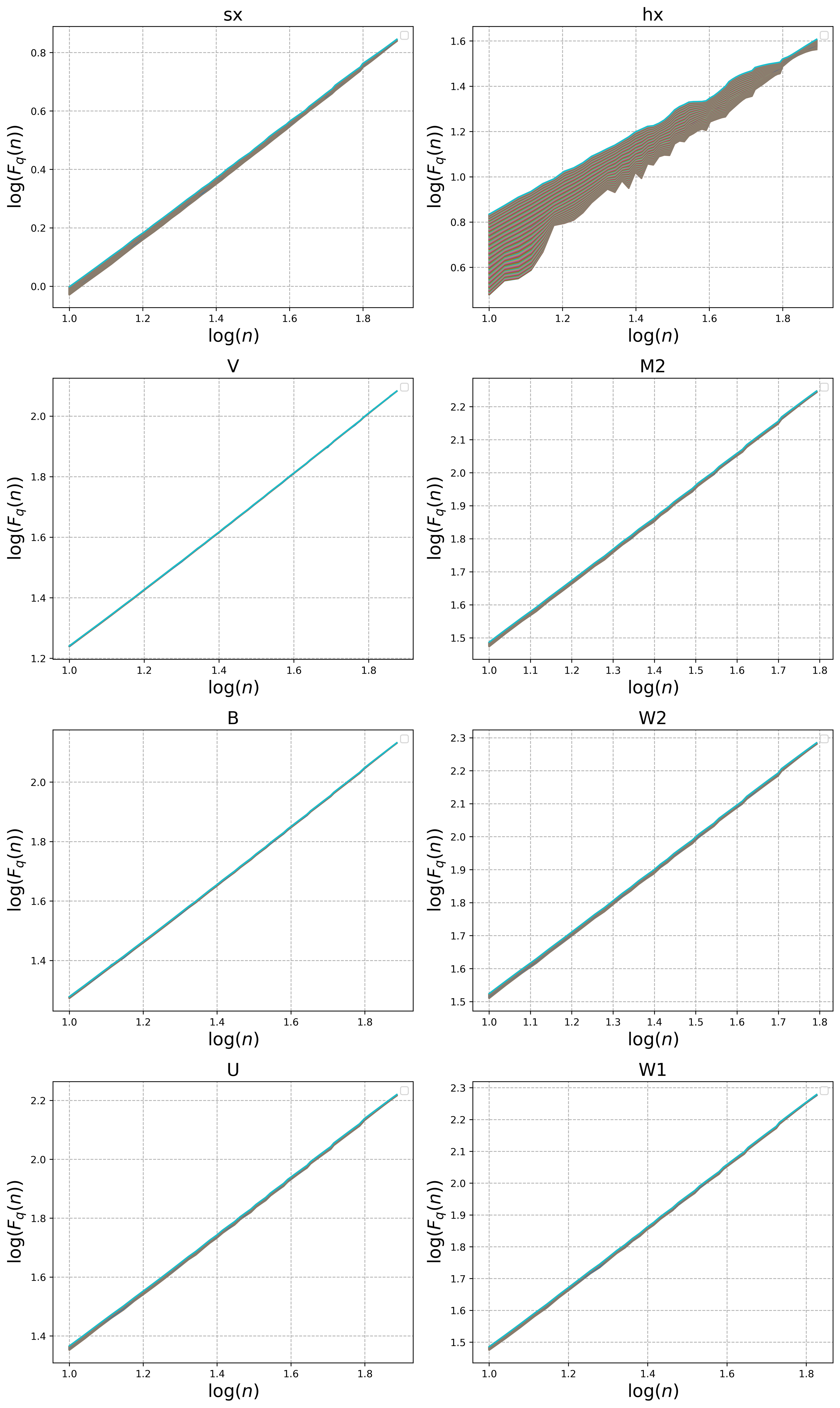}
\captionsetup{hypcap=false}\captionof{figure}{Fluctuation function $F_q(n)$ of the original light curve of NGC~4151 for all eight observational bands.}
\label{fig:fq_plot_3}

\clearpage

\noindent\includegraphics[width=\textwidth,height=0.80\textheight,keepaspectratio]{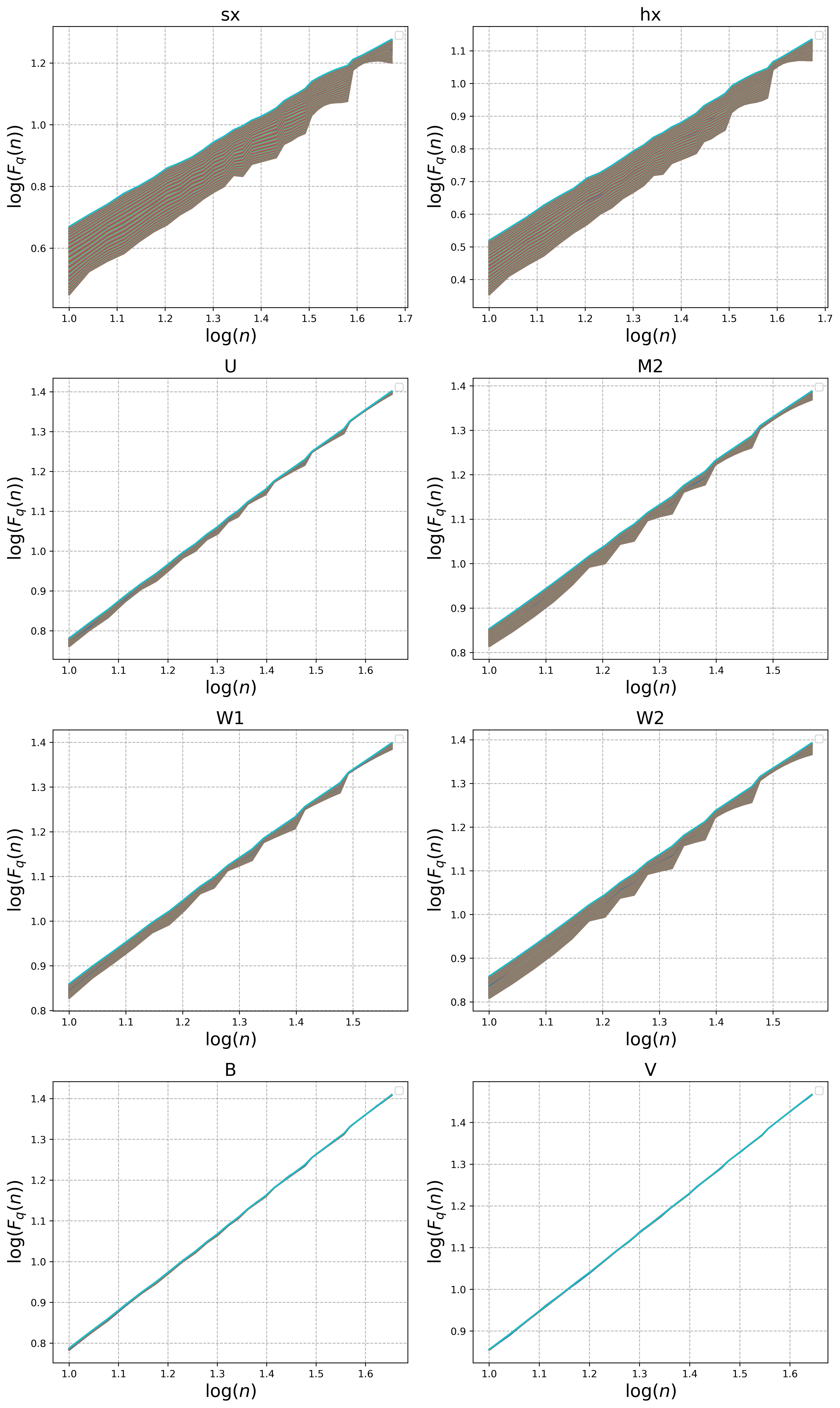}
\captionsetup{hypcap=false}\captionof{figure}{Fluctuation function $F_q(n)$ of the original light curve of NGC~4593 for all eight observational bands.}
\label{fig:fq_plot_4}
\par}
%%%%%%%%%%%%%%%%%%%%%%%%%%%%%%%%%%%%%%%%%%%%%%%%%%

% Don't change these lines
\bsp	% typesetting comment
\label{lastpage}
\end{document}